\documentclass[reprint, amsmath,amssymb,aps, prl,superscriptaddress]{revtex4-1}

\usepackage{amsfonts,amssymb}
\usepackage{graphics}
\usepackage{graphicx}
\usepackage{epsfig}
\usepackage{mathrsfs}
\usepackage{verbatim}
\usepackage{hyperref}    % links
\usepackage{bm}          % bold math letters
\usepackage{color}
\usepackage{ulem}

\begin{document}

\title{Photoluminescence in array of doped semiconductor  nanocrystals}

\author{K.V. Reich}
\email{Reich@mail.ioffe.ru}
\affiliation{Fine Theoretical Physics Institute, University of Minnesota, Minneapolis, MN 55455, USA}
\affiliation{Ioffe Physical-Technical Institute, St Petersburg, 194021, Russia}
\author{T. Chen}
\affiliation{Fine Theoretical Physics Institute, University of Minnesota, Minneapolis, MN 55455, USA}
\author{Al. L. Efros}
\affiliation{Naval Research Laboratory, Washington, D.C. 20375, USA}
\author{B.I. Shklovskii}
\affiliation{Fine Theoretical Physics Institute, University of Minnesota, Minneapolis, MN 55455, USA}

\begin{abstract}
We study the dependence of the quantum yield of photoluminescence of a dense, periodic array of semiconductor nanocrystals (NCs) on the level of doping and NC size. Electrons introduced to NCs via doping quench photoluminescence by the Auger process, so that practically only NCs without electrons contribute to the photoluminescence. Computer simulation and analytical theory are used to find a fraction of such empty NCs as a function of the average number of donors per NC and NC size. For an array of small spherical NCs, the quantization gap between $1S$ and $1P$ levels leads to transfer of electrons from NCs with large number of donors to those without donors. As a result, empty NCs become extinct, and photoluminescence is quenched abruptly at an average number of donors per NC close to $1.8$. The relative intensity of photoluminescence is shown to correlate with the type of hopping conductivity of an array of NCs.
\end{abstract}

%\pacs{63.70.+h, 63.20.-e, 05.60.Cd,64.70.qd,44.10.+i}
\date{\today}

\maketitle

Semiconductor nanocrystals (NCs) can be used as building blocks for new solid materials with bulk properties which do not exist in conventional solids. From monodisperse spherical NCs one could assemble a closely packed NC solid with a three-dimensional (3D) periodic structure with spacing between NCs much smaller than the NC diameter. ~\cite{Murray-AnnuRev30-2000,Vanmaekelbergh-ChemSocRev34-2005,Talapin-ChemRev110-2010,Hanrath-JVST30-2012}. Like in bulk semiconductors, doping is critical for NC solids, which would otherwise be electrically insulating. Adding charged
carriers dramatically increases the electronic mobility of a NC array  \cite{TalapinPbSe,LiuLaw-NL6-2010,Cordones-JACS134-2012,Guyot-Sionnest-JPCL3-2012, SahuNanoLett12} and opens new possibilities for many interesting applications, including field-effect transistors, light-emitting devices, photodetectors, and solar cells \cite{LeeTalapin-NatueNano6-2011,ChungTalapin-NL12-2012,Kagan-NanoLetter12,SunWise-NatureNano-2012,Shirasaki-NaturePhoto-2013}.

Due to intrinsic difficulties of NC doping by impurity atoms\cite{NorrisScience08}, the initial progress in the introduction of carriers into NCs was reached by putting the electron-donating molecules in the vicinity of the NC surface (remote doping) \cite{ShimNature00,YuGuyot-Sionnest-Science300-2003,TalapinSciencr05} and applying external electric fields (electrochemical doping) \cite{YuGuyot-Sionnest-Science300-2003,TalapinSciencr05,WangScience01,RoestPRL02}. Only recently was a successful demonstration of the carrier introduced into a NC via electronic impurity doping reported \cite{WangScience01,SahuNanoLett12,BrovelliNanoLett12,KangAdvMater13}. In order to improve transport properties of a 3D NC array by doping it with electronic impurity, at least the following three conditions must be satisfied. First, the intentionally introduced impurities should be shallow donors. Second, a donor electron should be delocalized within a NC by the confinement potential. This condition is fulfilled if the NC diameter $D$ is smaller than six Bohr radii of electrons, $a_B=\hbar^2 \kappa_{NC} /m e^2$, where $e$ is the electron charge, $\kappa_{NC} $ is the dielectric constant, and $m$ is the effective electron mass of a semiconductor ~\cite{Ekimov1990sad}. Third, the average number of deep surface states, which can trap an electron from the conduction band, should be significantly smaller than the average number of donors $\nu$ per NC. Currently, surface traps in NCs still play a dominating role. But the number of traps is constantly decreasing due to improved technology. For example, the use of inorganic ligands drastically reduces the number of surface states \cite{Zamkov_2013}. Therefore it is time to study what can be achieved in a trap-free case.

For this ideal situation, the transport properties of a NC array have been studied theoretically as a function of the average number of donors $\nu$ and the NC diameter\cite{Conductivity}. It turns out that doping results in transport properties, which are typical for disordered semiconductors. Indeed, the number of donors in different NCs is random. One could think that each NC has as many electrons as donors, so that all NCs remain neutral. It was  shown that in an array of small spherical NCs the quantization gap between the lowest $1S$ and $1P$ confinement levels leads to redistribution of electrons between NCs. Due to the energy difference between $1P$ and $1S$ levels,  electrons from a NC  with more than two donors, which in an isolated NC would occupy $1P$ levels, are transferred to NCs with zero or one donors, which have unfilled $1S$ levels. Thus, in the ground state of the NC array many NCs are charged.  The charged NCs lead to the random Coulomb potential  and to the Coulomb gap in the density of localized electron states that determines variable range hopping conductivity\cite{Conductivity}.

From an experimental point of view, it is interesting  whether the effect of electron redistribution can be detected by quantum yield (QY)  of photoluminescence (PL) of an array of NCs. It is known that the extra electrons placed in NCs by shallow donors  trigger the  nonradiative Auger  recombination of a photoexcited electron-hole pair. In such a process,  the annihilation energy of the electron-hole pair is transferred  to  an extra electron.  The rate of nonradiative Auger processes  is much larger than the rate of the radiative recombination \cite{FlitzanisAPL87, ChepicJL90, KlimovScience00, Efros_Auger_4}. Thus,  even one extra electron can quench photoluminescence from a NC with almost 100\% probability. In other words, only empty NCs, i.e. the ones without an electron, contribute to the photoluminescence.  A similar question of PL quenching by extra surface electrons induced by a gate is studied in Ref. \cite{Robel_Lee_Pietryga_Klimov_2013}, while here we concentrate on PL quenching by electrons in the bulk array of NCs.

In this  paper we calculated the fraction of empty NCs $f(\nu)$, which controls the PL quantum yield  as a function of the average number of donors $\nu$ per  NC and   the NC size. We show that  if $\nu$ reaches the critical value of $1.8$, there are practically no empty NCs in a NC array due to electron redistribution between NCs, and consequently,  PL vanishes almost completely.  Our calculations  also show that the relative intensity of  the PL  is  strongly correlated with  the transport  properties of these array. We have suggested an experiment which could provide important information about  the redistribution of the charges between NCs affecting both QY of PL of an array of NCs and its hopping conductivity.

If NCs are neutral, the number of empty NCs is equal to the number of NC with no donors. When donors are added randomly to each NC,  the probability $P(N)$ that a given NC has exactly $N$ donors is given by the Poisson distribution:
\begin{equation}
\label{eq:Poisson}
P(N) = \frac{\nu^N}{N!} \exp(-\nu).
\end{equation}
Therefore $f(\nu)$ can be obtained from Eq. \eqref{eq:Poisson} at $N=0$. This ``Poisson'' fraction is $f_P(\nu) = \exp(-\nu)$.  Below we show that $f(\nu)$ gives only the upper bound for  $f(\nu)$, so that, actually, $f(\nu) < f_P(\nu)$. This happens because electrons redistribute between NCs in order to minimize their total energy \cite{Conductivity}. 

In order to calculate the fraction of empty NCs, we assume that  NCs are identical spheres of diameter $D$ that form  a three dimensional cubic lattice structure.  We assume that the overlap of electron states of neighboring NCs is so small that a small disorder originating from small fluctuations of diameters $D$ and a random Coulomb potential (see below) leads to  Anderson localization of each electron in one of the NCs. As a result, each NC has exactly an integer number of electrons. At the same time, as we mentioned above, in the case $D < 6a_B$,  the wave function of a donor electron is delocalized within a NC \cite{Ekimov1990sad, Efros2000eso}. We also suppose that the wave function is close to zero at the NC surface due to large confining potential  barriers created by the insulator matrix surrounding NCs. Under these conditions the kinetic energy $E_Q(n)$ of the $n$th  electron added to a NC in the parabolic band approximation is :
 
\begin{equation}
E_Q(n) = \frac{\hbar^2}{m D^2} \times \left\{ 
\begin{array}{llr}
0, &  n = 0 & \\
19.74, &  n = 1, 2 & ~~~(1S)\\
40.38, &  3 \leq n \leq 8 & ~~~(1P)\\
.....\\
\end{array}
\right.
\label{eq:EQ}.  
\end{equation} 
As a result, the first two electrons added to a NC fill its $1S$ level, the next six ones fill its $1P$ level, and so on. 

The  kinetic energy of electrons is only a part of the total energy of a NC. One should add to it the total Coulomb energy of donors and electrons and their interaction. In general,  calculating the total Coulomb energy of the system is a difficult problem, since the positions of positive donors are random within the NC's volume.  For our problem, however, a significant simplification is available because the internal dielectric constant $\kappa_{NC}$ typically is much larger than both the external dielectric constant $\kappa_i$ of the insulator in which the NCs are embedded and the overall effective dielectric constant $\kappa$ of the array of NCs. Specifically, the large internal dielectric constant $\kappa_{NC}$ implies that any internal charge $q$ is essentially completely compensated by the NC dielectric response, which leads to  homogeneous redistribution of the charge major part, $q (\kappa_{NC} - \kappa)/\kappa_{NC}$, over the NC surface.  In this way each semiconductor NC can be considered as a metallic one in terms of its Coulomb interactions; namely, total Coulomb energy(self energy) is equal to $(N-n)e^2/\kappa D$. These approximations are equivalent to the so-called constant interaction model, which is commonly used for individual quantum dots \cite{Meir1991tts}.

For illustration, let us discuss why in our approximation one can neglect a donor position dependent correction to the kinetic energy $E_Q (N)$ from Coulomb interaction of an electron with a donor. This correction, by the order of magnitude, is equal to $e^2/\kappa_{NC} D$ \cite{Efros2000eso}. One can see that if $\kappa \ll \kappa_{NC}$, this correction is always smaller than charging energy and therefore much smaller than the $S$-$P$ gap energy, which induces redistribution of electrons. For more detailed discussion of the energy of the system, see the Appendix, where we show that the energy of a NC is completely determined by only two terms: quantum kinetic energy and self energy.
 
The  gap $\delta E=E_Q(3)-E_Q(2)$ between the $1S$ and $1P$ levels of a NC and the energy of adding one electron to  a neutral NC  $e^2/\kappa D$, which we call the charging energy, are the two most important energies of our theory. Below we use their ratio,
\begin{equation}
\Delta = \frac{\delta E}{e^2/\kappa D} = \frac{20.64 \kappa \hbar^2}{me^2D } = 20.64 \frac{\kappa}{\kappa_{NC}} \frac{a_B}{D},  
\label{parameter}
\end{equation}
which grows with decreasing NC diameter $D$. Let us estimate this parameter. Consider, for example,  CdSe NCs with $\kappa_{NC} \simeq 9.2$ (Ref. \cite{Madelung})  and $D = 5$ nm arranged in a crystalline array with lattice constant $D'=6$ nm. Let us assume that  $k_i=2$. The Maxwell-Garnett formula \cite{Maxwell,Doyle,Conductivity} then gives  $\kappa \simeq 3.2 \ll \kappa_{NC}$. Using $m \simeq 0.12 ~ m_e$ ($m_e$ is the electron mass) we have $a_B \simeq 4$ nm, $e^2/\kappa D \simeq 0.08$ eV and $\Delta \simeq 5.7$.

We see that for the case of CdSe our main approximation based on the inequality $\kappa \ll \kappa_{NC}$ has only 30\% accuracy. But this is the price for the relatively simple  solution of an otherwise intractable problem of calculating the total energy depending on positions of all donors in non-uniform dielectric constant media. Ignoring positions of donors, our solution preserves larger and more important effects of fluctuations of the number of donors in NCs \cite{Conductivity}.  For other semiconductors the inequality $\kappa \ll \kappa_{NC}$ can be much stronger. For example, for PbSe, $\kappa_{NC} \simeq 210$ \cite{Madelung}, and for the same $D$ and $D'$ one gets  $\kappa \simeq 4.5 \ll \kappa_{NC}$ resulting in 2\% accuracy.

We see from Eq. \eqref{parameter} that the  range of applicability of our model,  $D \lesssim 6 a_B $, for  CdSe NCs corresponds to $\Delta > 1$. In principle, one can consider an array of NCs with even smaller $\kappa/\kappa_{NC}$ and $D \simeq 6a_B$, where the ratio $\Delta < 1 $ . In that case, one can expect no effects of $\Delta$ on $f(\nu)$, so that $f(\nu) = f_P(\nu)$.

It is easy to see that at  large  $\Delta \gg 1$  electrons have to  be redistributed between NCs, charging originally neutral NCs . Imagine, for example, two  neutral NCs of the same size, one  with three donors and the other NC with no donors at all. In the $N=3$ NC, two electrons occupy the $1S$ level, and the third one occupies the $1P$ level. Obviously, in equilibrium the electron from the $1P$ level of the first NC will move to the  $1S$ level of the second NC. Even though this process charges both NCs, in the case $\Delta \gg 1$ the electron gains energy approximately equal to  $\delta E$. In other words, the first NC acts as a donor, and the second one acts as an acceptor.

In order to calculate $f(\nu)$  at any $\Delta$ and to take into account the Coulomb interaction between charged NCs, we model a 3D NC array with the   Hamiltonian \cite{Conductivity}
\begin{eqnarray}
H  & = &\sum_i \left[ \frac{e^2 (N_i - n_i)^2 }{\kappa D} + \sum_{k = 0}^{n_i}E_Q(k) \right] \nonumber \\
& & + \sum_{\langle i,j \rangle } \frac{e^2 (N_i - n_i)(N_j - n_j)}{\kappa r_{ij}},
\label{eq:H}
\end{eqnarray}
where $e(N_i - n_i)$ describes the charge of the $i$-th NC.  The first term of the  Hamiltonian is the sum of self-energies of NCs, the second term describes the total quantum energy of  $n_i$ electrons in the $i$th NC, and the last term is responsible for the Coulomb interaction between different NCs. The interaction between two NCs $i$ and $j$ at a distance $r_{ij}$ is  written as $q_i q_j/\kappa r_{ij}$. Such an approximation is obvious for a large distance $r_{ij}$. But one can show that even for two nearest neighbor NCs in a typical array, this approximation works well. For parameters $D = 5$ nm and $D’ = 6$ nm the energy of interaction of the two nearest charged metallic spheres can be approximated as the interaction of two central point charges $q_i q_j/\kappa r_{ij}$ with an accuracy of 1\% \footnote{To calculate the accuracy one can use exact result (1.12) from Ref. \cite{Lekner_2012} or estimate sixth term in the series (2.2) from \cite{Lekner_2012}}.  Of course, the same statement is applicable for two nearest neighbor NCs if $\kappa_{NC} \gg \kappa$.

Our goal is to find the fraction of empty NCs $f(\nu)$ in the ground state. Such a state is defined by the set of electron occupation numbers $\{n_i\}$ that minimize the Hamiltonian $H$. We simulate a finite, cubic array of $8000=20\times20\times20$ NCs and use the approach outlined in Ref. \cite{Conductivity} to calculate the ground state. First, we specify  $\nu$ and the total number of donors in our system $M=8000\nu$. Then we assign a random donor number $\{N_i\}$ for each NC $i$ according to Eq.\ (\ref{eq:Poisson}) and the condition $\sum_i N_i = M$.  The initial values of the electron numbers $\{n_i^0\}$ are then assigned randomly in such a way that the system is overall neutral, i.e., $\sum_i n_i^0 = M$. The program then searches for the ground state by looping over all NC pairs $\langle i j \rangle$ and attempting to move one electron from $i$ to $j$. If the move lowers the Hamiltonian $H$, then it is accepted; otherwise, it is rejected. In this way we arrive at a pseudoground state, which describes the NC array at low temperatures \cite{Conductivity}. Then, the number of empty NCs in the pseudoground state is averaged over 50 different realizations of the random initial arrays of donors numbers $\{N_i\}$ .  

\begin{figure}
\includegraphics[width=1\linewidth]{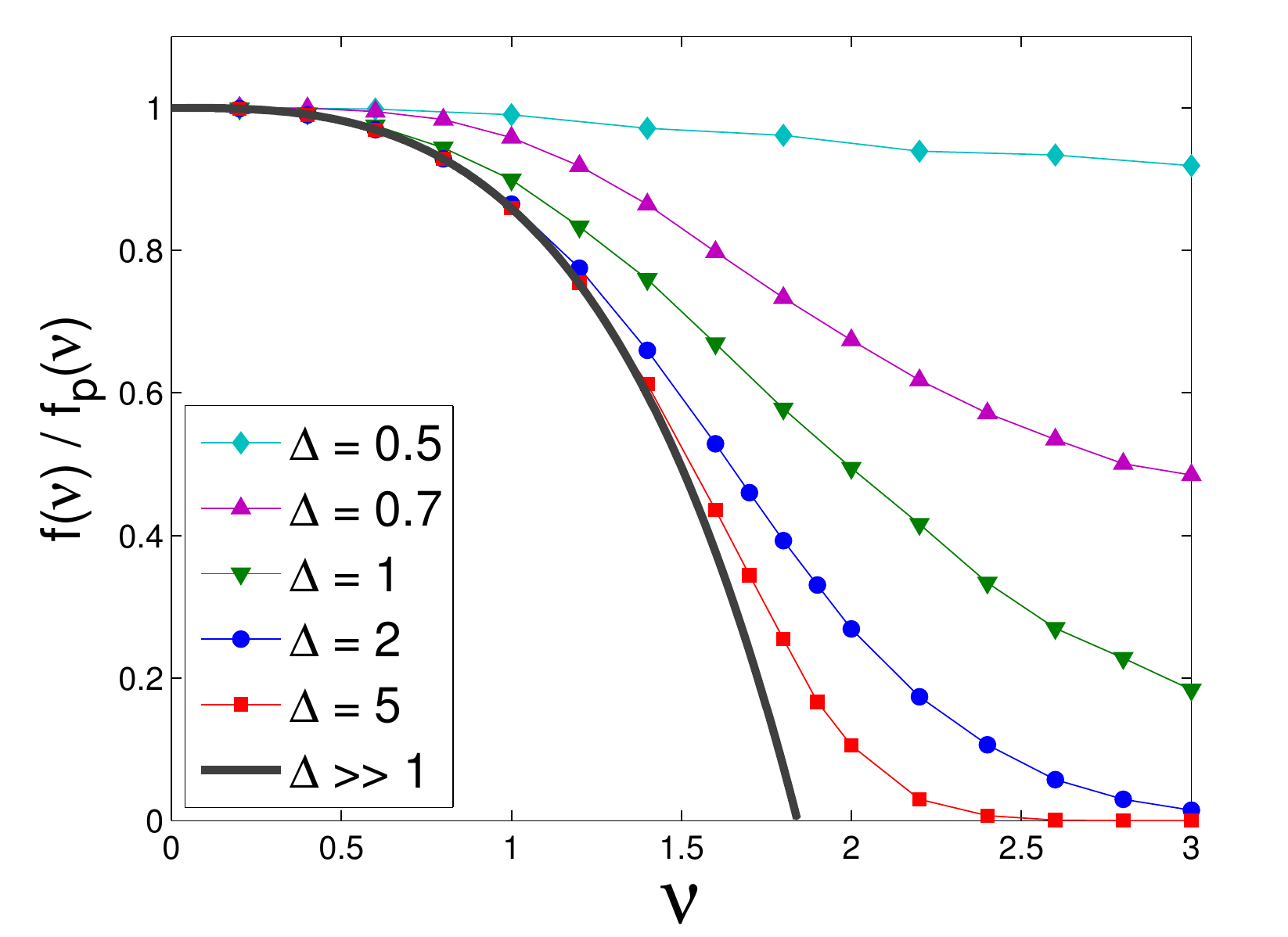}
\caption{(Color  online) 
The dependence of the fraction of empty NCs $f(\nu)$  in an  array of doped NCs on the average number of donors per NC $\nu$. The function $f(\nu)$ is normalized  to the fraction of  undoped NCs which is described by  the Poisson distribution: $f_P(\nu)=\exp(-\nu)$. The calculations were  conducted  for several  ratios of the $1S$-$1P$ gap  to the charging energy, $\Delta$. Numerical results are shown by  symbols connected by eye guiding thin lines. The size of the points reflects a calculation uncertainty. The analytical solution for the case $\Delta \gg 1$ is shown by the solid black line.}
\label{fig:empty}
\end{figure}

In Fig. \ref{fig:empty} we show the results of a numerical simulation for the  fraction of empty NCs  $f(\nu)$  for different  $\Delta$. Recall  that the fraction of empty donors is proportional to the QY of PL. We limit the plot to the range $\nu \le 3$ because, as we show below, at $\nu \gtrsim 3$ QY is already determined by filled NC. In the limit  $\nu \ll 1$, the system remains nearly uniformly neutral due to an extreme scarcity of NC with $N > 2$, which can provide electrons to $N=0,1$ NCs.  Indeed, according to Eq.\ \eqref{eq:Poisson}, at small $\nu$ the fraction of NC with $N > 2$ is $\simeq \nu^3/6 \ll 1$. The difference between $f(\nu)$ and $f_P(\nu)$ increases at $\nu>1$. For example, we see that for the case $\Delta=5$ the fraction $f(\nu)$ vanishes almost completely at $\nu>2$.   

In order to understand our numerical result let us find analytically $f(\nu)$ for the limit $\Delta \gg 1$  (using Hamiltonian $H$ without the last term of Eq.\ \eqref{eq:H} ). In this case, all the electrons from $1P$, $1D$ ...  levels of NCs with $N=3,4....$ (which we call excessive electrons) tend to occupy the $1S$  level in NCs with $N=0$ and 1. Let us start from small enough $\nu$ when all excessive electrons can still find places in 1S level in NCs with $N=0,1$.  As  shown above,   a NC with $N=3$ donors provides one excessive electron. In general, a NC with $N \ge 3$ provides $N-2$ excessive electrons. The total number of excessive electrons $S(\nu)$ per NC can be readily derived:

\begin{equation}
S(\nu)=\sum \limits_{N=2}^{\infty} (N-2) P(N) = \nu -2 +(\nu+2)\exp(-\nu). 
\label{eq:excess_electron}
\end{equation} 

Neutral NCs with $N=2$ have a filled $1S$ level and have no $1P$ electrons. Unoccupied $1S$ states are available only in NCs with $N=0$ and $N=1$, and excessive electrons could move to one of these NCs. (Recall that an $N=1$ NC has one electron before electrons redistribute between NCs). The  first states to fill are the two lowest energy states, namely, the first electron state in the $N=0$ NC and the second electron state in the $N=1$ NC. Indeed, according to Hamiltonian  \eqref{eq:H}, there is no energy difference between the first electron state in the $N=0$ NC and the second electron state in the $N=1$ NC, but bringing the second electron to an N=0 NC requires more energy. Thus, the first electron state in the $N=0$ NC and the second electron state in the $N=1$ NC share excessive electrons proportional to their fractions $\exp(-\nu)$ and $\nu \exp(-\nu)$ in the array. As a result, the probability to find an excessive electron in the $N=0$ NCs is proportional to the fraction of such NCs $1/(1+\nu)$. When $f(\nu)>0$, we arrive at
 
\begin{equation}
f(\nu)= \exp(-\nu) - \frac{1}{1+\nu} S(\nu) ~~~~ (\Delta \gg 1).
\label{eq:excess_electron}
\end{equation} 
Here, the first term is the fraction of the $N=0$, NCs and the second term describes the number of them lost due to filling by excessive electrons.

The function  $f(\nu)$ obtained for $\Delta \gg 1$ (Eq.\ \eqref{eq:excess_electron}) is shown in Fig. \ref{fig:empty} by the full line. One can see that $f(\nu)$ vanishes at $\nu = \nu_c \simeq 1.8$ and $f(\nu)=0$ at $\nu>\nu_c$. One can see that the  numerical result for $\Delta=5$ is close to  our analytical result for $\Delta \gg 1$. The long range random potential originating from the last term of Eq.\ \eqref{eq:H} and so far ignored leads to the avoided threshold in the vicinity $\nu= 1.8$. 
Curves for smaller $\Delta$ show how the $1P$-$1S$ gap-induced carrier redistribution weakens and $f(\nu)$ approaches $f_P(\nu)$.

In order to try to quantify the charge redistribution, we call arrays of NCs in which $f(\nu)>f_P(\nu)/2$ ``bright'', while we call NC arrays obeying the opposite inequality $f(\nu)<f_P(\nu)/2$  ``dark''. With this definition Fig. \ref{fig:empty} implies  that a  NC array is dark' for $\Delta \gtrsim 0.7$ and $\nu \gtrsim 1.5$ and is bright' otherwise. We would like to note a correlation between these results and the main phase diagram of the hopping transport \cite{Conductivity}. For the same model of the array of NCs,  it was shown in Ref. \cite{Conductivity} that the conductivity is described by the Efros-Shklovskii variable range hopping law for $\Delta \gtrsim 0.5 $  and  $\nu \gtrsim 0.7$  and the thermally  activated nearest-neighbor hopping law in other cases. Thus there is a correlation between QY and the conductivity of an array of NCs. One should expect the Efros-Shklovskii law in a dark' NC array, while the thermally-activated law should be expected in a bright' NC array. There are  also strong correlations  between the brightness   and  photoconductivity of the NC array. An extra electron that quenches the PL in a NC leaves that NC and can be involved in photoconductivity. Thus there is complementarity between PL and photoconductivity in arrays of NCs. When PL is quenched and a NC array is dark, the photoconductivity is strong and vice versa.

Until now we  have neglected completely  the radiative decay of charged NC,  assuming  that the  ratio, $R$, of the radiative recombination rate to the  Auger rate for a NC with one extra electron is very small, namely $R \ll 1$. 

\begin{figure}
\includegraphics[width=1\linewidth]{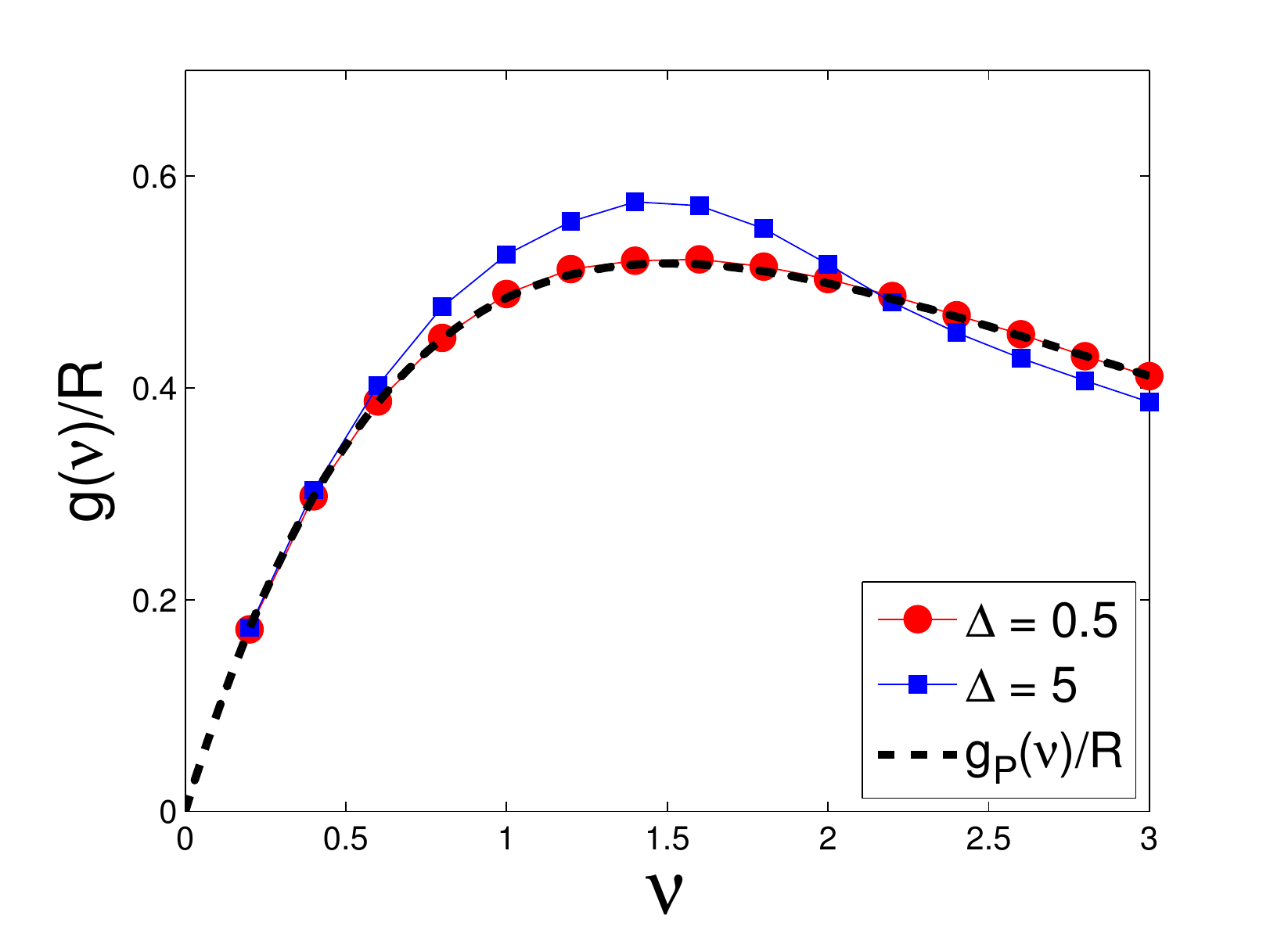}
\caption{(Color  online) The dependence of the   photoluminescence QY of all filled NCs $g(\nu)/R$ related to the QY of an undoped array, on the  average number of donors per NC $\nu$. The calculations were  conducted  for the  ratios of the $1S$-$1P$ gaps to the charging energies $\Delta=0.5$ and 5. Numerical results are shown by  circles and squares connected by eye guiding thin lines. The size of the circles and squares  reflects a calculation uncertainty. The analytical  dependence of  $g_P(\nu)/R$   calculated for $\Delta\ll 1$ according to Eq. \eqref{eq:g} is shown by the dashed line.}
\label{fig:fillled}
\end{figure}

At relatively large $\nu > 2 $ the fraction of empty NCs $f(\nu)$ becomes very small.  In that case, we have to take into account PL from NCs with electrons (filled NCs). We assume below that the rate of the Auger process linearly increases with the number of electrons in a NC, $n$, so that the QY from this NC is proportional to $R/n$. Thus, the normalized to undoped NC array QY of all filled NCs $g(\nu)$ is 

\begin{equation}
\label{eq:P}
  g(\nu) = R \sum \limits_{n=1}^{\infty} \frac{w(n)}{n}.
\end{equation}
Here,  $w(n)$ is the fraction of NCs with $n$ electrons in the NC array.  At small $\Delta$ there is no electron redistribution in the array of NCs, i.e. $n = N$, and we can use the  Poisson distribution  of Eq.\ \eqref{eq:Poisson} for $w(n)$. This gives
\begin{equation}
\label{eq:g}
 g_P(\nu) = R  e^{-\nu}\int \limits_0^{\nu} \frac{e^x-1}{x} dx,  
\end{equation}
The function  $g_p(\nu)$  is shown in Fig. \ref{fig:fillled} by the dashed line. One can see that in the range of $1< \nu  <3$ $g_P(\nu)$, is close to $0.5R$ .  

In order to study the role of  the redistribution of electrons at larger $\Delta$ we simulated the ground state of electrons in an array of NCs and computed the distribution function $w(n)$ averaged over ten realizations of a $20\times 20 \times 20$ NC array. Then we calculated $g( \nu)$ for $\Delta = 0.5$ and $\Delta = 5$ according to Eq.\ \eqref{eq:P}. These numerical results are shown in Fig. \ref{fig:fillled}. We found that at small $\Delta = 0.5$  our $g( \nu)$ practically coincides with $g_P(\nu)$, as expected.  Remarkably, our numerical result for $g(\nu)$ for $\Delta = 5$  does not differ from Eq.\ \eqref{eq:g} by more than by $20 \%$. 
One can understand this in the following way.
At small $\nu < 1$ the fraction of NC with $N>2$ is small; there is no electron redistribution between NCs regardless of the value of $\Delta$, and therefore $g(\nu) \simeq g_P(\nu) $. At $ 1 \lesssim  \nu \lesssim  3$,  many NCs  have two  electrons either because they have two donors or because of  electron redistribution between $N=3$ NCs and $N=1$ NCs. In that case we can estimate QY of PL from these NCs as $R/2$. Thus for any $\Delta$ one can say that the contribution of filled NCs is roughly $g(\nu)=\alpha R$, where $\alpha$ is close to $0.5$ in the range of $1 < \nu  < 3$ . To find at what $\nu$ all the filled NCs begin to contribute to QY more than empty ones we  equate $f(\nu)$ and $\alpha R$. For example, at $\Delta = 5$ and $R=0.01$ this happens at $\nu \approx 2.3$, while at $R=0.1$ filled NCs win at $\nu \approx 1.7 $.

Although all  the numerical and analytical results of this paper are derived  under the assumption that temperature $T=0$, our theory is applicable for any reasonable temperature. The theoretical  restrictions  of our theory on the temperature are $k_BT < e^2/\kappa D $, which means  $T < 800$\,K ,  for example, in the case of  an array of CdSe NCs with $D=5$ nm, for which $ e^2/\kappa D \simeq 0.08$\, eV. 

We also neglect completely the dispersion of NC size. Our theory, however, can be used also in the NC array  with dispersion of   NC size if the typical fluctuation of the ground state energy is smaller than the charging  energy.  In the closely packed  array of NCs this dispersion could not be higher than 5\% because the latter would prevent formation of the ordered NC  array \cite{Murray-AnnuRev30-2000}.  The  5\% fluctuation of the NC sizes results in the fluctuation of the  $1S$-$1P$ gap $\delta E$, whose  magnitude can be estimated as $\sim 0.1 \delta E$ in the parabolic band approximation.  For a such NC array this means $0.1 \delta E < e^2/\kappa D$, or $\Delta < 10$.   The last condition is satisfied in practically all interesting cases.

 One needs to note that PL could also be quenched  by  the  Foster resonant energy transfer (FRET) of the exciton from an empty NC to   neighboring  NCs with electrons, where this exciton undergoes the fast  non-radiative Auger recombination \cite{Schins_Houtepen_Siebbeles_2013}.  This requires the FRET time  to be shorter than  the exciton radiative recombination time.   The exciton cannot be transferred to a NC which has more than one electron because  significant additional  energy equal the 1S-1P gap  is  required for such a transition. Indeed, the third electron could occupy only the $1P$ electron level.  This energy restriction  does not exist for NCs with one electron, and at a first glance, the FRET could be as efficient  as  in  transitions between empty NCs.  The analysis of the FRET between the empty NCs and  NCs with one electron shows, however, that this is not the  case because the fast non-radiative Auger decay rate $\gamma_A$ of an exciton in NCs with one electron suppresses the FRET rate.  The  FRET rate in the last case is equal to  $ 1/\tau_{FRET}=(2\pi/\hbar^2)W^2/\gamma_A$, where $W$ is the  Foster  matrix element that describes coupling between the exciton state in empty NCs and  NCs with one electron  \cite{Agranovich} .  One can see that FRET time, which  for empty NCs is on the order 3--10\,ns, is actually much longer than this value because $W/\hbar\gamma_A \ll 1$.

The theory we developed  deals with  spherical NCs with a non-degenerate parabolic electron spectrum, which exist, for example, in  CdSe NCs. In  NCs made of semiconductors with several degenerate conduction band minima, for example, PbS, PbSe, and Ge with four degenerate ellipsoids, the $1S$-level accommodates $4 \times 2 = 8$ electrons. The redistribution of electrons due to the 1S-1P gap at $\nu$ = 8 does not affect QY because the fraction of  NCs with $N=0$ at such $\nu$ is practically equal to zero (recall that $f(\nu) < \exp(-\nu)$). However, in PbS, PbSe, and Ge NCs, the $1S$-level is split by the surface potential into two levels with degeneracies of 2 and 6 \cite{degenerate}. One can therefore use our theory to describe PL in the range of $\nu < 3$ in PbS, PbSe, and Ge NCs using this smaller gap. If this gap results in $\Delta < 1$, we arrive at $f(\nu) = f_P (\nu)$. 

The  gap value  is always critical for the carrier redistribution. In large spherical NCs made of non-degenerate single-band semiconductors the 1S-1P gap decreases faster with growing size than the charging energy. This leads to $\Delta < 1$, and  we again end up with  $f(\nu) = f_P (\nu)$.  The effective gap can be significantly  reduced even in small NCs, whose shape substantially deviates from the spherical. The splitting of the $1P$ levels in such structures creates several substantially narrower gaps instead of one big one, which results in $f(\nu) = f_P (\nu)$. 

We would like to suggest an experiment which could allow one to study the carrier redistribution in an array of doped NCs. Namely, the time dependence of the PL intensity after the short pulse excitation should allow  measurement of the hopping time between NCs in a NC array. In equilibrium the fraction of empty NC is determined by $f(\nu)$. Photo-excitation of NCs with electrons triggers the non-radiative Auger processes, which ionize NCs, creating excited, almost free electrons\cite{Shabaev2013}. These electrons are captured back by NCs randomly. This fast process leads to the redistribution of carriers and increases the number of empty NCs. Their fraction again is determined by Poisson distribution $f_P(\nu)$,  which in turn increases the PL intensity. The relaxation to the equilibrium distribution function $f(\nu)$ is much slower and is controlled by the hopping time between NCs with three electrons and NCs with no electrons. The effect could be measured by the pump probe experimental technique in a NC array with low mobility, where the hopping time is much longer than the exciton decay time. The first short, intense pump pulse initially creates a non-equilibrium distribution of empty NCs $f_P(\nu)$. The relaxation of the electrons to the equilibrium distribution $f(\nu)$ could be measured with the help of the time dependence of the PL intensity excited by the probe pulse.

In conclusion, this paper  has shown that in an array of small NCs due to the quantization gap induced redistribution of electrons among NCs, empty NCs responsible for PL become extinct at an average number of donors per NC equal to 1.8, leading to abrupt quenching of PL.

\begin{acknowledgments}
The authors would like to thank B. Skinner and D. Talapin for helpful discussions. This work was supported primarily by the MRSEC Program of the National Science Foundation under Award Number DMR-0819885 and T. Chen  was partially supported by the FTPI.  Al.L.E. acknowledges  the  financial support  of  the Office  of  Naval Research (ONR) through the Naval Research Laboratory Basic Research  Program. 
\end{acknowledgments}

\appendix* 

\section{Appendix: Derivation of total energy for one NC}

We consider redistribution of electrons between NCs with different numbers of donors. For that we need to write down the full Hamiltonian for all NCs. Here we calculate total energy $E_{N,n}$ of a NC with $N$ donors and $n$ electrons. First, we consider the energy of an electron and a donor in a neutral NC in more detail. In order to do this we consider a sphere of radius $R=D/2$, and the dielectric coefficient $\kappa_{NC}$ is surrounded by a medium of coefficient $\kappa$.  In the one electron approximation the total energy $E_{1,1}$ of the system can be determine from the  Schrodinger equation:
$$
\left(-\frac{\hbar^2}{2m} \nabla ^2   + U(r_1,r_2) \right) \Psi = E_{1,1} \Psi(r_1)
$$ 
where $\Psi(r_1)$ is the wave function of the electron with mass $m$.   $U(r_1,r_2)$ is the energy necessary to assemble the donor  and  the electron  in positions $r_1$ and $r_2$ inside the sphere. This energy can be exactly written as \cite{Brus_1984}:
$$U = - \frac{e^2}{\kappa_{NC}|\boldsymbol{r_1}-\boldsymbol{r_2}|} - U_M(r_1,r_2) + U_c(r_1)+U_c(r_2),$$
where the first term describes the screened Coulomb interaction of the electron and the donor. $U_M$ is a mutual polarization term that can be thought of as the interaction of one charge with  the surface polarisation charge created by the other charge. $U_c$ describes an interaction between the charge and its own image arising due to the charge polarization on the surface of the sphere.

$$
U_M = \frac{e^2}{R} \sum \limits_{l=0}^\infty \alpha_l  \frac{r_1^lr_2^l}{R^{2l}}  P_l(\cos(\theta))
$$

where $\theta$ - is the angle between $\boldsymbol{r_1}$ and $\boldsymbol{r_2}$,$P_l$ is Legendre polynomial,  and $\alpha_l=   (\kappa_{NC}-\kappa)  (l+1)/\kappa_{NC}(l\kappa_{NC}+(l+1)\kappa)$

$$
U_c(r)=\frac{e^2}{2 R} \sum \limits_{l=0}^\infty \alpha_l \left(\frac{r}{R}\right)^{2l}
$$

In the main approximation, when  $\kappa \ll \kappa_{NC}$, these formulas become simpler:

$U_M = e^2/\kappa R$, $U_c= e^2/2 \kappa R$ and

$$U = \frac{e^2}{2 R \kappa } + \frac{e^2}{2 R \kappa } - \frac{e^2}{\kappa  R} = 0
$$

In that case, the total energy of system is $E_{1,1}=E_Q$, where $-\hbar^2/2m \nabla^2 \Psi = E_Q\Psi$.

One can generalize this result for the case with  $n$ electrons and $N$ donors in the NC. Again, in the case where $\kappa \ll \kappa_{NC}$, $U$ doesn't depend on electron and donor position and can be written as total self-energy $(N-n)^2e^2/2 R \kappa$. The total energy for electrons and donors can be written in the form $E_{N,n}=\sum_i E_Q(i) + U = \sum_i E_Q(i) + (N-n)^2e^2/2 R \kappa$. This can be seen as separate kinetic energy for all electrons  and self energy of NCs with the charge $N-n$. 

%

%\bibliography{Photoluminescence_in_QD}
\end{document}